# Main-sequence Turnoff Stars as Probes of the Ancient Galactic Relic: Chemo-dynamical Analysis of a Pilot Sample

Renjing Xie[1,2], Haining Li[1], Ruizhi Zhang[1], Yin Wu[1,2], Xiang-Xiang Xue[1], Gang Zhao[1,2], Shi-Lin Zhang[1], and Xiao-Jin Xie[1,2]

[1] CAS Key Laboratory of Optical Astronomy, National Astronomical Observatories, Chinese Academy of Sciences, Beijing 100101, People's Republic of China; lhn@nao.cas.cn
[2] School of Astronomy and Space Science, University of Chinese Academy of Sciences, No.19(A) Yuquan Road, Shijingshan District, Beijing, 100049, People's Republic of China

## Abstract

The main-sequence turnoff (MSTO) stars well preserve the chemical properties where they were born, making them ideal tracers for studying the stellar population. We perform a detailed chemo-dynamical analysis on moderately metal-poor ($-2.0 <$ [Fe/H] $< -1.0$) MSTO stars to explore the early accretion history of the Milky Way. Our sample includes four stars observed with high-resolution spectroscopy using ESPaDOnS at the Canada–France–Hawaii Telescope and 163 nearby MSTO stars selected from the SAGA database with high-resolution results. Within the action-angle spaces, we identified Gaia–Sausage–Enceladus (GSE, 35 objects), stars born in the Milky Way (in situ, 31 objects), and other substructures (21 objects). We find that both GSE and in situ stars present a similar Li plateau around $A$(Li) $\sim 2.17$. GSE shows a clear $\alpha$-knee feature in Mg at [Fe/H] $\sim -1.60 \pm 0.06$, while the $\alpha$-elements of in situ stars remain nearly constant within this metallicity range. The iron-peak elements show little difference between GSE and in situ stars except for Zn and Ni, which decrease in GSE at [Fe/H] $> -1.6$, while they remain constant for in situ stars. Among heavy elements, GSE shows overall enhancement in Eu, with [Ba/Eu] increasing with the metallicity, while this ratio remains almost constant for in situ stars, suggesting the contribution of longer timescale sources to the s-process in GSE. Moreover, for the first time, we present the r-process abundance pattern for an extremely r-process enhanced (r-II) GSE star, which appears consistent with the solar r-process pattern except for Pr. Further investigation of larger GSE samples using high-resolution spectra is required to explore the reason for the significantly higher Pr in the GSE r-II star.

*Unified Astronomy Thesaurus concepts:* Milky Way dynamics (1051); Chemical abundances (224); Milky Way stellar halo (1060)

## 1. Introduction

The ΛCDM cosmological model predicts that the Milky Way formed hierarchically through accreting and merging with dwarf galaxies, which left abundant substructures in the Galactic halo (S. D. M. White & M. J. Rees 1978; S. D. M. White & C. S. Frenk 1991; S. Cole et al. 2000; J. S. Bullock et al. 2001; A. Zolotov et al. 2009). The discovery of Sagittarius streams in the position–velocity phase space has provided the first piece of evidence of this scenario in the Milky Way (R. A. Ibata et al. 1994). The dynamics, chemical abundances, and ages of stars preserve important information about the environments in which they were born, allowing us to trace their origins in chemo-dynamical spaces. Searching for these substructures and understanding their origins are crucial for us to replay the formation and evolution history of the Milky Way.

While integrals of motion (IoMs) have been commonly used for identifying accretion relics (e.g., A. Helmi et al. 1999; G. C. Myeong et al. 2019), recent numerical simulations demonstrate that single accretion events can fragment into multiple phase-space substructures over time, undermining the long-term IoM coherence (e.g., I. Jean-Baptiste et al. 2017; G. Pagnini et al. 2023). Neither dynamics nor chemistry alone can fully disentangle the Galactic halo's assembly history. Therefore, a combined chemo-dynamical approach is essential.

The Gaia satellite provides magnificent astrometry for billions of stars and radial velocities (RVs) for tens of millions of stars (Gaia Collaboration et al. 2023; D. Katz et al. 2023), enabling us to study high-precision, all-sky 6D dynamical characterization of Milky Way stars. The combination of the Gaia data and the large spectroscopic surveys, such as Radial Velocity Experiment (RAVE; M. Steinmetz et al. 2006), Large Sky Area Multi-Object Fiber Spectroscopic Telescope (LAMOST; G. Zhao et al. 2006, 2012), Galactic Archaeology with HERMES (GALAH; G. M. De Silva et al. 2015), Apache Point Observatory Galactic Evolution Experiment (APOGEE; S. R. Majewski et al. 2017), H3 Survey (C. Conroy et al. 2019), etc., allows us to gather accurate 6D phase space and chemical information from large samples of halo stars and helps us to obtain a much clearer picture of how the Milky Way, especially its older components, has evolved since ∼10 Gyr ago. Using these data, dozens of substructures in the Galactic stellar halo have been identified, including Gaia–Sausage–Enceladus (GSE; V. Belokurov et al. 2018; M. Haywood et al. 2018; A. Helmi et al. 2018), Thamnos (H. H. Koppelman et al. 2019), Sequoia (G. C. Myeong et al. 2019), and Wukong/LMS-1 (R. P. Naidu et al. 2020; Z. Yuan et al. 2020).

To further constrain the origin of these overdense regions, information about their chemical abundances is needed, which is available from observations of their moderate-to-high-resolution spectra. The chemical abundances of the substructure contain information about star formation, evolution, and their interaction with the interstellar medium. Different groups of elements can be synthesized through different types of nucleosynthesis processes in different astrophysical sites, and thus different types of nucleosynthesis events are characterized by their unique chemical signatures. For example, the ratio of $\alpha$-elements to iron ([$\alpha$/Fe]) is commonly used to trace the star formation history in a stellar system. The $\alpha$-knee refers to the observed phenomenon that the abundance ratio of [$\alpha$/Fe] starts to decrease when the stellar system evolves to a certain metallicity, indicating the onset of type Ia supernovae (SNe Ia) in galaxies (e.g., F. Matteucci & E. Brocato 1990; F. Matteucci 2003). The rapid neutron-capture process (r-process) is responsible for synthesizing about half of the isotopes of elements heavier than iron in the Universe (A. Frebel & A. P. Ji 2023). Stars displaying relatively prominent enhancements in heavy elements are regarded as key objects for constraining the r-process (A. Frebel 2018). Recent works have studied different components in the Galactic halo, combining chemistry and dynamics (e.g., T. Matsuno et al. 2019, 2021, 2022a, 2022b; S. Monty et al. 2020; D. S. Aguado et al. 2021; D. Horta et al. 2023; R. Zhang et al. 2024).

The most prominent Galactic substructure known to date in the Galactic halo is the GSE (V. Belokurov et al. 2018; M. Haywood et al. 2018; A. Helmi et al. 2018), which was first discovered by P. E. Nissen & W. J. Schuster (2010, 2011, 2012) as the low-$\alpha$ sequence. P. E. Nissen & W. J. Schuster (2010, 2011, 2012) argued that the lower-$\alpha$ sequence corresponds to populations accreted from dwarf galaxies. This prediction was subsequently confirmed by data from the Gaia satellite and was identified as the Gaia–Sausage and Gaia–Enceladus by V. Belokurov et al. (2018) and A. Helmi et al. (2018), respectively. The simulation results show that GSE is the relic of a nearly head-on major merger event $\sim$10 Gyr ago and dominates the nearby halo (M. Haywood et al. 2018; A. Helmi et al. 2018; A. Helmi 2020). Quite a number of efforts have been devoted to investigating the chemical features of GSE (D. S. Aguado et al. 2021; T. Matsuno et al. 2021; A. Carrillo et al. 2022; X. Ou et al. 2024; R. Zhang et al. 2024).

Moreover, exploring a proper metallicity range, e.g., where nucleosynthesis sources with different timescales could be separated, is very helpful and efficient in understanding their origins. The abundance ratios observed in all dwarf spheroidal galaxies for stars on the metal-poor side of the knee tend to be indistinguishable from those in the Milky Way halo (E. Tolstoy et al. 2009), and the Galactic substructures in the very metal-poor region ([Fe/H] $< -2.0$) would not show significant differences in most elements from the in situ components (R. Zhang et al. 2024). The most significant difference occurs within the metallicity range from $-2.0$ to $-1.0$, which is best distinguished by chemistry from one stellar system to another (see, e.g., T. Matsuno et al. 2022a; D. Horta et al. 2023). The peaks of the metallicity distribution function of most substructures are also in this range (A. Helmi et al. 2018; G. C. Myeong et al. 2019, 2022; R. P. Naidu et al. 2020). Thus, a combination of high-precision kinematics and detailed chemical abundances of a statistical sample in the $-2.0$ to $-1.0$ metallicity range can better reflect the properties of the progenitor systems of these accretion events, allowing us to better understand the formation and evolution of the Galaxy.

Chemo-dynamical analysis holds the key to unraveling early nucleosynthesis and the origins of diverse elements. Previous studies of Galactic substructures including the GSE have mainly used giant stars as tracers (D. S. Aguado et al. 2021; T. Matsuno et al. 2021; A. Carrillo et al. 2022; G. Limberg et al. 2024; X. Ou et al. 2024). The advantages of giants include being relatively bright and able to trace relatively distant halo, and the spectral lines of giants are easier to measure. However, the surface elemental abundances of giants may be affected by stellar evolution. Both the main sequence (MS) and the main-sequence turnoff (MSTO) represent the stages before the first dredge-up, allowing for a better study of the original properties of progenitor systems. MSTO stars are relatively brighter than MS stars, which allows effective data to be gathered during limited observing time Therefore, MSTO stars are ideal tracers for studying the chemical abundances of stellar populations (A. D. Mackey et al. 2008; P. Goudfrooij et al. 2009; S. Zhang et al. 2019; M. Xiang & H.-W. Rix 2022).

We hereby conduct a study to explore the early accretion history of the Milky Way through moderately metal-poor ($-2.0 \leqslant$ [Fe/H] $\leqslant -1.0$) MSTO stars selected from the LAMOST (G. Zhao et al. 2006, 2012), which has obtained more than 10 million low-resolution spectra by LAMOST DR5. Here, we present the results of a pilot sample of moderately metal-poor MSTO stars combining a supplement sample of SAGA database data. This project aims to investigate the properties of substructures as presented by MSTO stars. This paper is organized in the following way. In Section 2.1, we briefly introduce the program sample of our work and describe the techniques used to obtain the chemical abundances. Section 2.2 introduces the analysis of the literature sample, and Section 2.3 presents the dynamical parameters of the final sample. The results and discussions are presented in Section 3. Finally, in Section 4, we draw the main conclusions of this study.

## 2. Observations and Analysis of Sample

### 2.1. Program Sample: LAMOST and CFHT Observations

Candidates for the program sample were first selected from the LAMOST DR5, whose spectra cover the wavelength range from 3700 to 9100 Å with resolution ($R$) $\sim$ 1800 (G. Zhao et al. 2006, 2012), allowing for reliable estimation of stellar parameters in the metallicity range we focused on. The preliminary candidates for moderately metal-poor MSTO stars have been selected by adopting the following criteria:

(i) $-2.0 <$ [Fe/H] $< -1.0$,
(ii) $3.0 < \log g < 5.0$,
(iii) 5500 K $< T_{\rm eff} <$ 7000 K,
(iv) Signal-to-noise ratio (S/N) $>$ 50 in $g$ band.

As a pilot project, we further selected the brightest targets with $V < 11$, for which high-resolution spectroscopy with sufficient quality could be obtained within a limited observation time. This selection resulted in 45 MSTO candidates from the LAMOST DR5, as shown in Figure 1. Five of them were observed in 2020 November, with the high-resolution spectrograph ESPaDOnS at the Canada–France–Hawaii Telescope (CFHT; J. F. Donati et al. 2006) using the "star" mode. However, one of these targets was found to be in a binary system, so we ultimately retained four targets for further study.

Table 1
Observing Runs

| Object ID | Dates of Observation | Total Exposure Time | S/N (4500 Å) |
|---|---|---|---|
| LAMOST J072231.45+084913.0 | 2020 Nov 27 | 1200 s | 110 |
| LAMOST J091220.95+203227.7 | 2020 Nov 27 | 1300 s | 117 |
| LAMOST J091733.95+472839.6 | 2020 Nov 27 | 1300 s | 120 |
| LAMOST J093101.03−030405.3 | 2020 Nov 28 | 1800 s | 136 |

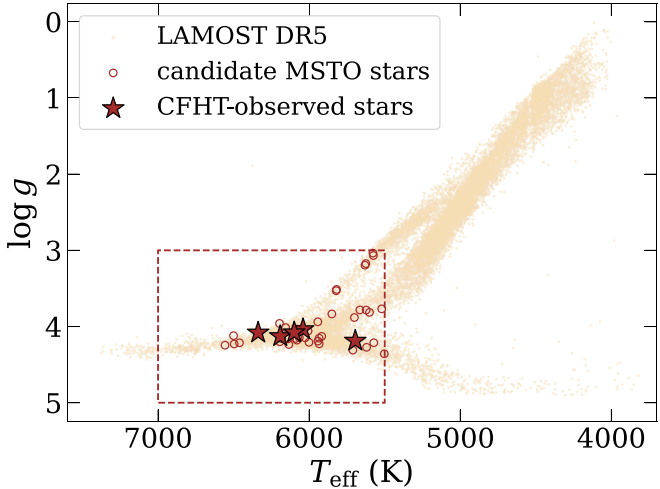

**Figure 1.** A Hertzsprung–Russell diagram of selected and observed MSTO stars in LAMOST DR5. The orange dots are all moderately metal-poor stars with $g$-band S/N > 50 in LAMOST DR5. The brown open circles are MSTO candidates, and the brown stars represent five objects observed with CFHT. The dashed brown rectangle is the criteria we adopted to select MSTO stars.

The detailed observing information of the four program stars is shown in Table 1.

The observation obtained high-resolution ($R = 68{,}000$) spectra from 3600 to 10800 Å, with exposure times ranging from ∼20 to 30 minutes. The data reduction was performed with the pipeline of ESpRIT,[3] including bias subtraction, flat fielding, wavelength calibration, and spectral extraction.

After that, RVs were measured using the fxcor task in the Image Reduction and Analysis Facility (IRAF; D. Tody 1993) determined by cross correlation with the template spectrum of a metal-poor star. The typical errors of the RV measurements from the CFHT high-resolution spectra are $\lesssim 1$ km s$^{-1}$. The RVs, as determined from the CFHT high-resolution spectra, are presented in Table 2. For all four targets, there are multiple observations with LAMOST. Within the typical uncertainty of LAMOST RV measurement (10 km s$^{-1}$, W. Aoki et al. 2022), we did not find any significant variation in their RVs, and hence no clear evidence of binary systems based on current observation has been found for the program star.

### 2.1.1. Equivalent Width Measurements

We used an IDL code Tool for Automatic Measurement of Equivalent width (TAME; W. Kang & S.-G. Lee 2012) to fit a Gaussian profile for the corresponding single and unblended atomic absorption lines. We measured the equivalent widths (EWs) of the atomic lines for all program stars, with the linelist referring to L. Mashonkina et al. (2010), W. Aoki et al. (2013), I. U. Roederer et al. (2022), and H. Li et al. (2022).

---
[3] https://www.cfht.hawaii.edu/Instruments/Spectroscopy/Espadons/Espadons_esprit.html

One star in our program sample has been studied by E. M. Holmbeck et al. (2020), which is thus treated as a calibration star for EW measurements. Figure 2 shows a comparison between our EW measurements and those of E. M. Holmbeck et al. (2020). The offset between the EW measurements is negligible, demonstrating the agreement between the reduced spectra and our EW measurements. The scatter is typically less than 0.30 mÅ, which is comparable to the typical uncertainty of EW measurements.

### 2.1.2. Stellar Parameters

For program stars, we used the infrared flux method and parallax data to determine the effective temperature and surface gravity, respectively. We also derived the stellar parameters with the spectroscopic method for verification and comparison. The atmospheric parameters were derived through an iterative procedure.

We derived the photometric effective temperature through the infrared flux method for further analysis because, on one hand, the nonlocal thermodynamic equilibrium (NLTE) effects for low-metallicity stars are significant (K. Lind et al. 2011; A. M. Amarsi et al. 2016); on the other hand, the extinction of our program stars is negligible (∼0). The infrared flux method used the $(V - K)_0$ calibration from I. Ramírez & J. Meléndez (2005), where the $V$ magnitudes were obtained from AAVSO Photometric All-Sky Survey (APASS; A. Henden & U. Munari 2014) and $K$ magnitudes were transformed from 2MASS $K_s$ (M. F. Skrutskie et al. 2006). The surface gravity was calculated using the online code PARAM 1.3 (L. da Silva et al. 2006), based on the relationships between parallax, bolometric flux, temperature, mass, and $\log g$.

Based on the derived $T_{\rm eff}$ and $\log g$, we calculated metallicity using Fe I and Fe II lines. Since the Fe abundances derived from Fe I lines can suffer from a relatively large NLTE effect, the average abundance obtained from Fe II lines is adopted as the final value of the metallicity. The microturbulent velocity ($\xi$) was derived to ensure no trend in Fe abundances from individual Fe I lines as a function of line strength. The tuning procedure was repeated until a reasonable convergence between ionization balance and metallicity was achieved.

For comparison and double checking, we also determined the stellar parameters using the spectroscopic method. We derived $T_{\rm eff}$ from the excitation balance of Fe I lines and the $\log g$ by adjusting the $\log g$ to ensure that the Fe abundances derived from Fe I and Fe II lines are consistent. We analyzed approximately 200 isolated Fe I lines, whose EWs ranged from 10 to 175 mÅ within the wavelength range of 4062 to 6546 Å. The average difference between the two methods is 18 K for $T_{\rm eff}$ and 0.22 dex for $\log g$. These values are consistent with each other within the measurement uncertainty. Due to the reasons previously mentioned, we adopt the photometric set of parameters as our final values, as listed in the $T_{\rm eff}$ (pho) and $\log g$ (para) columns of Table 2, respectively.

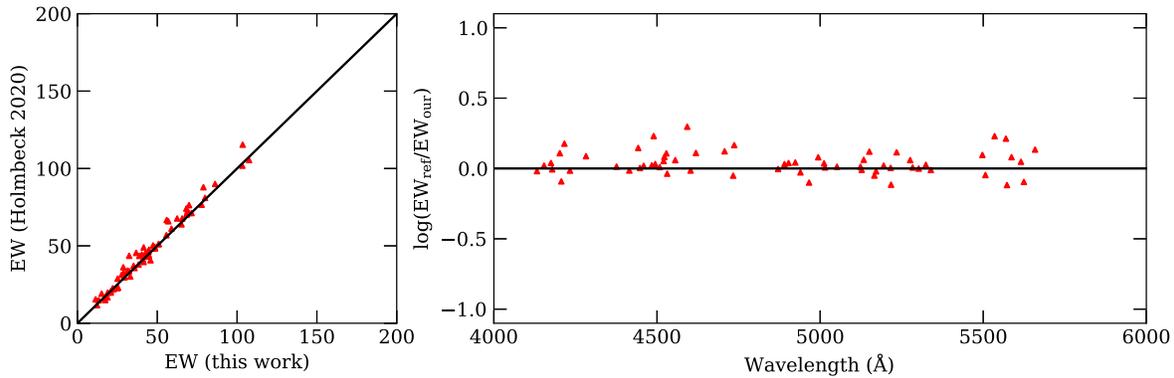

**Figure 2.** Comparison of EWs in this work with those in E. M. Holmbeck et al. (2020).

**Table 2**
Basic Information and Stellar Parameters of the Program Stars

| | Basic Information | | | | Adopted Parameters | | | | Spectroscopic Method | |
|---|---|---|---|---|---|---|---|---|---|---|
| ID | R.A. (deg) | Decl. (deg) | RV (km s$^{-1}$) | S/N | [Fe/H] | $T_{\rm eff}$(pho) (K) | log $g$(para) | $\xi$ (km s$^{-1}$) | $T_{\rm eff}$(spec) (K) | log $g$(spec) |
| J0722+0849 | 110.63 | +8.81 | −38.13 ± 0.10 | 110 | −1.72 ± 0.09 | 5938 ± 26 | 3.800 ± 0.051 | 1.35 ± 0.05 | 5975 ± 44 | 3.69 ± 0.05 |
| J0912+2032 | 138.08 | +20.54 | 209.71 ± 0.32 | 117 | −1.85 ± 0.08 | 5510 ± 19 | 4.654 ± 0.027 | 1.01 ± 0.13 | 5480 ± 48 | 4.66 ± 0.08 |
| J0917+4728 | 139.39 | +47.47 | −160.71 ± 0.67 | 120 | −1.86 ± 0.09 | 6204 ± 28 | 4.380 ± 0.053 | 1.45 ± 0.11 | 6080 ± 61 | 3.83 ± 0.09 |
| J0931−0304 | 142.75 | −3.06 | 124.91 ± 0.11 | 136 | −1.51 ± 0.09 | 6061 ± 28 | 3.830 ± 0.113 | 1.33 ± 0.10 | 6227 ± 51 | 4.02 ± 0.05 |

### 2.1.3. Elemental Abundances

We have determined chemical abundances for 27 elements, including Li, C, O, α-elements (Mg, Si, Ca, and Ti), light odd-$z$ elements (Na, Sc), iron-group elements (V, Cr, Mn, Fe, Co, Ni, and Zn), and neutron-capture elements (Sr, Y, Zr, Ba, La, Ce, Pr, Nd, Sm, Eu, and Gd). For all program stars, the S/N is higher than 100 around 5000 Å, which assures reliable measurements of abundances.

For the abundance analysis of all elements, the 1D plane-parallel, hydrostatic model atmospheres of the ATLAS NEW-ODF grid of F. Castelli & R. L. Kurucz (2003) were adopted, assuming a mixing-length parameter of $\alpha_{\rm MLT} = 1.25$, no convective overshooting, and local thermodynamic equilibrium (LTE). We used an updated version of the abundance analysis code MOOG (C. A. Sneden 1973), which treats continuous scattering as a source function that sums both absorption and scattering (J. S. Sobeck et al. 2011).

If the elemental abundance is derived from a single line, or different lines reflected abundance deviation larger than 0.2, the abundances can be examined with spectral synthesis. For blended lines, we used spectral synthesis results instead of EW abundances. Therefore, Li, C, V, Sr, Zr, Ba, La, Pr, Nd, Sm, Eu, and Gd abundances of the program stars were derived by computing synthetic spectra. Examples of the spectral fitting are shown for Li and C in Figures 3 and 4, respectively. In addition, hyperfine structure and isotopic splitting have been considered for Sr, Ba, La, and Eu (G. Borghs et al. 1983; A. McWilliam et al. 1995; J. E. Lawler et al. 2001a, 2001b). An example of the spectral fitting for Ba is presented in Figure 5.

The uncertainties in abundances are attributable to errors associated with the stellar atmospheric parameters and line-to-line scatter which is caused by line blending and continuum normalization. The abundance error caused by the stellar parameters was derived by the variation of the four atmospheric parameters within 1σ confidence intervals. The error due to the uncertainty of measurement was estimated from the standard deviation ($\sigma_X$) of abundances derived from individual lines and the number of lines ($N_X$), as $\sigma_X/\sqrt{N_X}$. This estimate provides a kind of random error also including the uncertainties of $gf$ values of spectral lines. When the number of effective lines measured for the element was less than 4, the larger $\sigma_X/\sqrt{N_X}$ and $\sigma({\rm Fe~I})/\sqrt{N_X}$ were then adopted.

To simplify the evaluation of the systematic errors associated with the stellar atmospheric parameters, we assumed that their uncertainties are independent, following the methodology adopted in previous studies (B. Cseh et al. 2018; M. P. Roriz et al. 2021). Table 3 provides the obtained abundances or upper limits, the total error derived by adding individual errors in the quadrature is given in the column of $\sigma_{\rm total}$, together with the number of lines used to derive the abundance. The photospheric solar abundances of M. Asplund et al. (2009) were adopted when calculating the abundance ratios ([X/Fe]).

### 2.2. Literature Sample: SAGA Database Compilation

Given the limited number of stars in our program sample for statistical discussions, we expanded the sample by selecting approximately 160 stars from the literature. The abundances of these stars were extracted from the SAGA database (S. Yamada et al. 2013; T. Suda et al. 2017). The SAGA database compiles abundances of metal-poor stars from studies that employed high- or medium-resolution spectrographs ($R \gtrsim 7000$) and effectively provides chemical abundances of many elements for a large number of metal-poor stars.

Since the SAGA database is based on abundance data collected from various studies, major sources of abundance uncertainties in the literature sample arise from different methods of abundance analyses, such as the selection of absorption lines. However, these uncertainties do not significantly affect our discussion of the chemical and kinematic properties since they behave similarly in stars with similar metallicities and temperatures (see, e.g., T. Matsuno et al. 2019, which is exactly the case for this study).

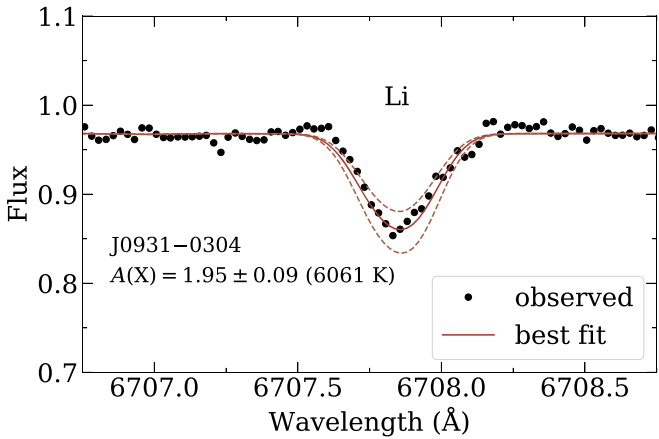

**Figure 3.** Example of spectral fitting of Li 6707.8 Å line for J0931 under LTE models. The observed spectrum is shown with dots. The best-fitting results of $A(X) = 1.95$ and uncertainty ($\pm 0.09$) have been shown in solid and dashed lines, respectively.

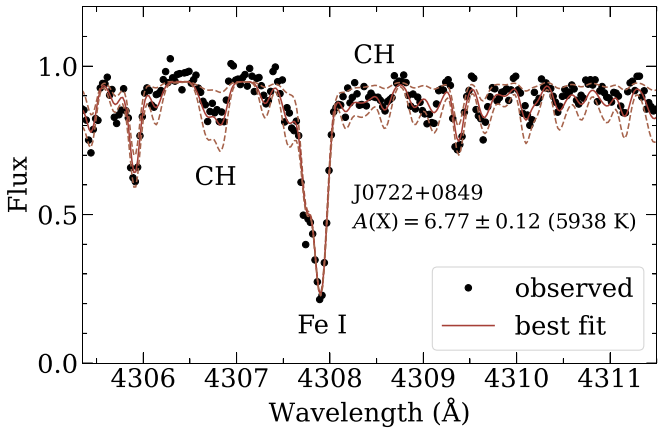

**Figure 4.** Best fits (continuous curve) of the CH features near 4310 Å for J0722. The meanings of the dots and lines are the same as in Figure 3, with the best-fitting result of $A(X) = 6.77$ and with uncertainty of $\pm 0.12$.

To ensure the element abundance analysis remains comparable in accuracy, we made use of only high-resolution results from the SAGA database. When multiple measurements existed for the same star, we prioritized those from studies that provided results for more stars, ensuring better consistency within our sample. In other cases, we chose data from studies providing more elements. Finally, we constructed the literature sample of 164 MSTO stars by chemical selection.

### 2.3. Determination of Dynamical Parameters

In order to associate stars with ancient merger events, we calculated the IoMs of the selected stars. Since the sample stars are located near the Galactic plane, we adopted an axisymmetric gravitational potential and computed the energy and actions, which are IoMs in axisymmetric systems.

We cross matched both the program sample and the literature sample with the Gaia DR3 catalogs (Gaia Collaboration et al. 2023) and the distance catalog of C. A. L. Bailer-Jones et al. (2021). We removed stars with relative distance uncertainties of C. A. L. Bailer-Jones et al. (2021) greater than 20% or Gaia

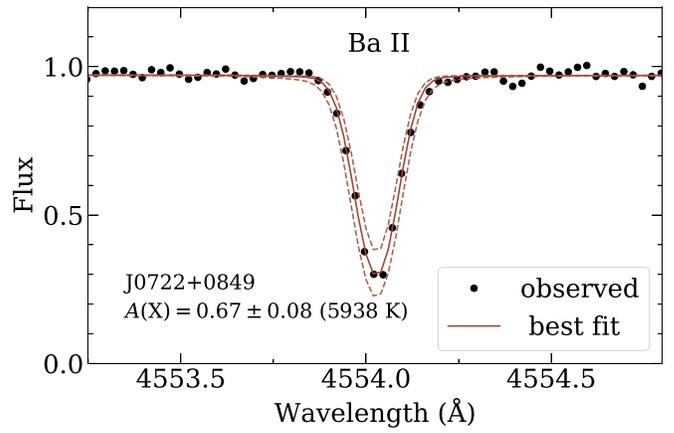

**Figure 5.** Example of spectral fitting of Ba 4554 Å line for J0722 under LTE models. The meanings of the dots and lines are the same as in Figure 3, with the best-fitting result of $A(X) = 0.67$ and with uncertainty of $\pm 0.08$.

Renormalised Unit Weight Error (RUWE) greater than 1.4 to ensure the astrometric measurements are reliable (L. Lindegren et al. 2021). Given that the Gaia RUWE is known to be sensitive to the presence of unresolved companions, the relatively small RUWE ($<1.4$) of the four program stars also indicates less possibility of binarity (V. Belokurov et al. 2020; K. G. Stassun & G. Torres 2021).

Finally, all four program sample stars and 163 stars in the literature sample were selected for further analysis. We computed the positions and velocities of the stars adopting the photogeometric distance from C. A. L. Bailer-Jones et al. (2021), coordinates and proper motions from the Gaia DR3, and the RVs were derived from CFHT spectra and Gaia DR3 for program sample and literature sample, respectively. We assumed that the Sun is 20.8 pc above the Galactic plane (M. Bennett & J. Bovy 2019) and located at 8.21 kpc from the Galactic center (P. J. McMillan 2017). The velocities of the stars are corrected adopting $(U_\odot, V_\odot, W_\odot) = (11.1, 12.24, 7.25)$ km s$^{-1}$ (R. Schönrich et al. 2010) for the solar peculiar motion, and $V_{LSR} = 233.1$ km s$^{-1}$ (P. J. McMillan 2017) for the motion of the local standard of rest (LSR). We used a right-handed Galactocentric Cartesian coordinate system $(X, Y, Z)$, in which the Sun is located at $(-8.21, 0, 0.0208)$ kpc. We also used a right-handed Galactocentric cylindrical system $(R, \phi, z)$, which indicates the negative $v_\phi$ for stars with prograde orbits.

Using the software Agama (E. Vasiliev 2019), we further computed orbital energy and action with an axisymmetric potential model of the Milky Way (P. J. McMillan 2017). We ran a 10,000 times Monte Carlo (MC) simulation for each star and took the median value and half the difference between the 16th and 84th quantiles to represent the value and the uncertainty of each quantity.

It has been demonstrated that the best way to identify stars belonging to discrete merger events, such as GSE, is using their IoMs since their IoMs retain the signature of their progenitors over long periods (A. Helmi et al. 1999; G. C. Myeong et al. 2018; D. K. Feuillet et al. 2020). In this work, we used the radial action $\sqrt{J_R} > 30$ kpc$^{1/2}$ km$^{1/2}$ s$^{-1/2}$ and the $z$-component of angular momentum $|L_z| < 500$ kpc km s$^{-1}$ to select GSE, since this $J_R - L_z$ plane enables a relatively pure selection of GSE stars (D. K. Feuillet et al. 2020; T. Matsuno et al. 2021; A. Carrillo et al. 2024). To classify in situ stars, we took the criterion from T. Matsuno et al. (2021, $L_z > 0$ kpc km s$^{-1}$;

Table 3
Abundances and Abundance Error Estimations for the Program Star

| | J0722+0849 | | | | J0912+2032 | | | | J0917+4728 | | | | J0931−0304 | | | |
|---|---|---|---|---|---|---|---|---|---|---|---|---|---|---|---|---|
| | $\log\epsilon(X)$ | [X/Fe] | $\sigma_{total}$ | N | $\log\epsilon(x)$ | [X/Fe] | $\sigma_{total}$ | N | $\log\epsilon(X)$ | [X/Fe] | $\sigma_{total}$ | N | $\log\epsilon(X)$ | [X/Fe] | $\sigma_{total}$ | N |
| Li | 2.10 | ⋯ | 0.10 | 1 | 1.85 | ⋯ | 0.06 | 1 | 2.15 | ⋯ | 0.08 | 1 | 1.95 | ⋯ | 0.09 | 1 |
| C | 6.77 | 0.00 | 0.12 | 1 | 6.39 | −0.20 | 0.10 | 1 | 7.09 | 0.30 | 0.10 | 1 | 7.09 | 0.23 | 0.11 | 1 |
| O | 7.80 | 0.77 | 0.08 | 3 | 7.45 | 0.60 | 0.07 | 2 | 7.83 | 0.78 | 0.06 | 3 | 7.87 | 0.75 | 0.06 | 3 |
| Na I | 4.58 | 0.00 | 0.07 | 3 | 4.78 | 0.38 | 0.13 | 3 | 4.57 | −0.03 | 0.06 | 2 | 4.61 | −0.06 | 0.07 | 2 |
| Mg I | 6.11 | 0.17 | 0.08 | 3 | 5.87 | 0.11 | 0.12 | 4 | 6.19 | 0.23 | 0.11 | 5 | 6.31 | 0.28 | 0.06 | 4 |
| Si I | 6.07 | 0.22 | 0.07 | 2 | 5.77 | 0.10 | 0.025 | 4 | 6.14 | 0.27 | 0.10 | 11 | 6.09 | 0.15 | 0.13 | 9 |
| Si II | 6.16 | 0.31 | 0.23 | 2 | ⋯ | ⋯ | ⋯ | ⋯ | 6.13 | 0.26 | 0.05 | 1 | 6.02 | 0.04 | 0.09 | 2 |
| Ca I | 4.95 | 0.27 | 0.08 | 25 | 4.55 | 0.05 | 0.10 | 25 | 4.91 | 0.21 | 0.07 | 25 | 5.06 | 0.29 | 0.07 | 22 |
| Sc II | 1.58 | 0.09 | 0.12 | 14 | 1.18 | −0.13 | 0.11 | 8 | 1.58 | 0.07 | 0.07 | 11 | 1.61 | 0.03 | 0.05 | 10 |
| Ti I | 3.50 | 0.21 | 0.13 | 19 | 3.25 | 0.14 | 0.09 | 19 | 3.54 | 0.23 | 0.08 | 16 | 3.58 | 0.20 | 0.07 | 16 |
| Ti II | 3.62 | 0.33 | 0.14 | 28 | 3.34 | 0.23 | 0.09 | 30 | 3.66 | 0.35 | 0.09 | 28 | 3.71 | 0.33 | 0.09 | 27 |
| V I | 2.42 | 0.15 | 0.10 | 1 | 1.93 | −0.16 | 0.10 | 1 | 2.29 | 0.00 | 0.06 | 1 | 2.61 | 0.21 | 0.10 | 1 |
| V II | 2.44 | 0.17 | 0.17 | 3 | 2.46 | 0.37 | 0.11 | 2 | 2.65 | 0.36 | 0.07 | 4 | 2.66 | 0.26 | 0.11 | 4 |
| Cr I | 3.72 | −0.26 | 0.09 | 12 | 3.42 | −0.18 | 0.11 | 12 | 3.77 | −0.23 | 0.09 | 10 | 3.84 | −0.23 | 0.07 | 12 |
| Cr II | 4.02 | 0.04 | 0.11 | 3 | 3.94 | 0.01 | 0.07 | 1 | 4.10 | 0.01 | 0.05 | 3 | 3.96 | −0.11 | 0.08 | 4 |
| Mn I | 3.29 | −0.48 | 0.07 | 2 | 2.94 | −0.60 | 0.08 | 3 | 3.26 | −0.52 | 0.05 | 3 | 3.35 | −0.51 | 0.12 | 3 |
| Fe I | 5.78 | −0.06 | 0.09 | 147 | 5.65 | −0.01 | 0.09 | 122 | 5.75 | −0.11 | 0.09 | 179 | 5.88 | −0.05 | 0.09 | 141 |
| Fe II | 5.84 | 0.00 | 0.03 | 15 | 5.66 | 0.00 | 0.03 | 16 | 5.86 | 0.00 | 0.03 | 20 | 5.93 | 0.00 | 0.03 | 20 |
| Co I | 3.26 | −0.07 | 0.07 | 3 | 3.13 | −0.02 | 0.12 | 3 | 3.46 | 0.11 | 0.17 | 2 | 3.46 | 0.04 | 0.13 | 2 |
| Ni I | 4.38 | −0.18 | 0.09 | 14 | 4.05 | −0.13 | 0.08 | 11 | 4.48 | −0.10 | 0.08 | 11 | 4.49 | −0.16 | 0.08 | 12 |
| Zn I | 2.83 | −0.07 | 0.08 | 2 | 2.66 | −0.06 | 0.10 | 2 | 3.03 | 0.11 | 0.07 | 2 | 2.97 | −0.02 | 0.07 | 2 |
| Sr II | 1.65 | 0.44 | 0.15 | 2 | 1.03 | 0.00 | 0.11 | 2 | 1.73 | 0.50 | 0.09 | 2 | 1.60 | 0.30 | 0.07 | 2 |
| Y II | 0.45 | −0.10 | 0.07 | 4 | 0.06 | −0.31 | 0.08 | 2 | 0.83 | 0.26 | 0.06 | 3 | 0.71 | 0.07 | 0.13 | 3 |
| Zr II | 1.31 | 0.39 | 0.10 | 2 | 0.71 | −0.03 | 0.09 | 1 | 1.46 | 0.52 | 0.19 | 2 | 1.54 | 0.53 | 0.15 | 2 |
| Ba II | 0.67 | 0.15 | 0.08 | 3 | 0.34 | 0.00 | 0.16 | 3 | 0.74 | 0.20 | 0.08 | 2 | 0.71 | 0.10 | 0.07 | 3 |
| La II | −0.16 | 0.40 | 0.15 | 4 | −0.28 | 0.46 | 0.07 | 3 | 0.09 | 0.63 | 0.12 | 5 | 0.14 | 0.61 | 0.12 | 3 |
| Ce II | 0.09 | 0.17 | 0.07 | 2 | ⋯ | ⋯ | ⋯ | ⋯ | 0.90 | 0.96 | 0.12 | 3 | 0.63 | 0.62 | 0.10 | 7 |
| Pr II | 0.36 | 1.30 | 0.09 | 2 | −0.06 | 1.06 | 0.09 | 1 | ⋯ | ⋯ | ⋯ | ⋯ | ⋯ | ⋯ | ⋯ | ⋯ |
| Nd I | ⋯ | ⋯ | ⋯ | ⋯ | 0.61 | 1.03 | 0.05 | 3 | <0.61 | 0.72 | 0.16 | 3 | 0.43 | 0.58 | 0.05 | 2 |
| Nd II | 0.19 | 0.43 | 0.08 | 5 | 0.86 | 1.28 | 0.17 | 2 | <0.67 | 0.72 | 0.16 | 2 | 0.35 | 0.50 | 0.09 | 2 |
| Sm II | <−0.01 | 0.69 | ⋯ | 2 | ⋯ | ⋯ | ⋯ | ⋯ | ⋯ | ⋯ | ⋯ | ⋯ | <0.93 | <1.54 | 0.09 | 1 |
| Eu II | −0.39 | 0.75 | 0.07 | 2 | −1.07 | 0.25 | 0.08 | 1 | ⋯ | ⋯ | ⋯ | ⋯ | −1.05 | 0.00 | 0.10 | 1 |
| Gd II | 0.25 | 0.84 | 0.12 | 1 | ⋯ | ⋯ | ⋯ | ⋯ | ⋯ | ⋯ | ⋯ | ⋯ | ⋯ | ⋯ | ⋯ | ⋯ |
| Dy II | ⋯ | ⋯ | ⋯ | ⋯ | 0.84 | 1.58 | 0.05 | 2 | ⋯ | ⋯ | ⋯ | ⋯ | ⋯ | ⋯ | ⋯ | ⋯ |

**Note.** $\log\epsilon(X) = \log(N_X/N_H) + 12 = [X/H] + \log\epsilon_\odot(X)$.

$\sqrt{J_R} < 15$ kpc$^{1/2}$ km$^{1/2}$ s$^{-1/2}$) and used $\sqrt{J_z} < 20$ kpc$^{1/2}$ km$^{1/2}$ s$^{-1/2}$ to avoid contamination of Helmi streams (see, e.g., Figure 5 in R. Zhang et al. 2024). Adopting the above criteria, we identified 31 GSE stars and 35 in situ stars in our sample. We note that one star in the program sample has $\sqrt{J_R} \sim 18$ kpc$^{1/2}$ km$^{1/2}$ s$^{-1/2}$, which does not meet the criteria we adopted for in situ stars. However, due to the planar orbital properties ($z_{max} \sim 0.5$ kpc) and the in situ chemical features, we treat this star as in situ in this work.

Additionally, we identified members from other substructures, such as Thamnos (13), Sequoia (4), Helmi streams (3), and Wukong/LMS-1 (1), following the criteria from the literature (see Table C1 of R. Zhang et al. 2024, and references therein). The associated results are shown in Figure 6. To ensure a statistical sample for the study, we only discuss the chemical distribution of GSE and in situ components in the following discussions, based on the final sample of 31 GSE stars and 35 in situ stars.

## 3. Result and Discussion

In this section, we present the chemical abundance analysis of the final sample from Li, C, and O, to other groups of elements ($\alpha$-, light odd-Z, iron-peak, and heavy), respectively. Chemical properties of stars from substructures (mostly in GSE in this work) are compared with those born in the Galaxy, i.e., in situ stars.

### 3.1. Lithium

Lithium is regarded as a crucial component in constraining our understanding of the Big Bang Nucleosynthesis, as well as low-mass star formation and evolution in the early Milky Way. For warm, metal-poor dwarf stars, there is a nearly constant level of Li abundance across a wide range of metallicities (−3.0 < [Fe/H] < −1.0), which is known as Spite Plateau (see, e.g., F. Spite & M. Spite 1982; P. Molaro et al. 1997; S. G. Ryan et al. 1999; M. Asplund et al. 2006; J. Meléndez et al. 2010; L. Sbordone et al. 2010).

We obtained the Li abundance in 40 sample stars, including nine GSE stars and nine in situ stars. Figure 7 shows the relation between Li abundance and metallicity in our sample. We can see that the Li abundance of both the GSE and in situ stars shows a plateau in the metal-poor range (−2.0 < [Fe/H] < −1.0). The mean Li abundance is $A(Li) = 2.17 \pm 0.14$ for GSE members and $2.18 \pm 0.15$ for in situ stars, respectively. The GSE members and in situ stars show no significant difference in Li abundance or its scatter, consistent with previous studies (P. E. Nissen & W. J. Schuster 2010; G. Cescutti et al. 2020; J. D. Simpson et al. 2021). This supports the hypothesis that the formation environment of stars does not play an important role in the formation of the Spite Plateau.

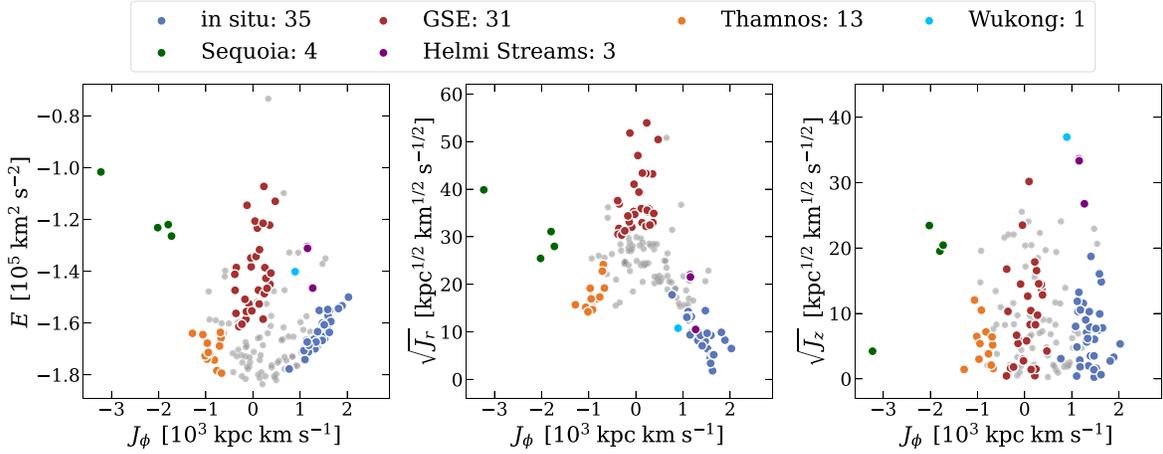

**Figure 6.** Kinematics of the stars in our sample. Different colors represent different structures: GSE (red), Thamnos (orange), Sequoia (green), Helmi streams (purple), Wukong/LMS-1 (sky blue), and in situ components (blue). The gray symbols mark stars that are not associated with the above substructures.

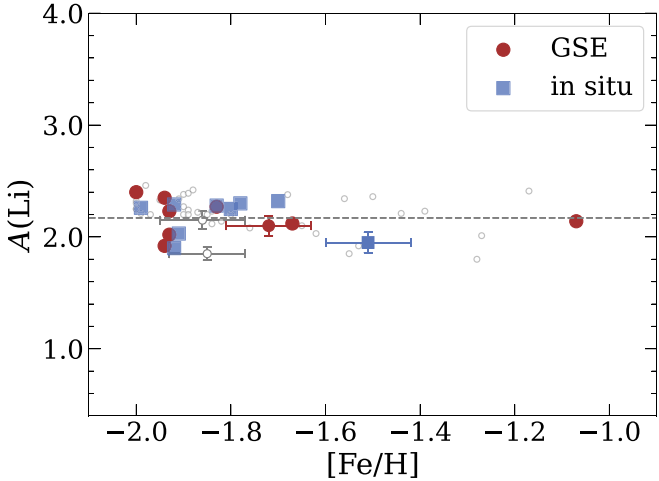

**Figure 7.** Li abundance of stars as a function of [Fe/H]. Symbols with and without error bars are stars from CFHT observation and the SAGA database, respectively. Different colors represent different origins: GSE stars (red), in situ components (blue), and stars not assigned to GSE or in situ components (gray). The gray dashed line represents $A$(Li) $\sim$ 2.17, which is the average Li abundance for GSE members.

### 3.2. Carbon and Oxygen

The carbon abundance in MSTO stars provides constraints on chemical evolution models (H. Li et al. 2022). Oxygen provides insights into the enrichment history of the interstellar medium, which is essential to deciphering the history of star formation in the Milky Way and its satellite galaxies (P. E. Nissen et al. 2014).

Carbon abundances were determined via spectrum synthesis for the CH band at 4305–4315 Å. The oxygen abundance was derived from the O I triplet at 7771, 7774, and 7775 Å. For eight GSE members and 12 in situ stars, carbon abundances could be measured, while nine GSE members and 11 in situ stars have oxygen measurements. There are only four stars having measurements of nitrogen abundance in our sample, so we will not discuss it here.

Figure 8 illustrates the [C/Fe] and [O/Fe] for the sample stars. Therefore, we do not need to introduce the carbon correction caused by stellar evolution. Metal-poor stars with carbon abundance enhancement are considered carbon-enhanced metal-poor (CEMP) stars (W. Aoki et al. 2007; J. Yoon et al. 2016, 2019). In this paper, we applied the criterion of W. Aoki et al. (2007). For MSTO stars, stars with [C/Fe] $> 0.7$ are considered as CEMP. CEMP stars can be subsequently classified into CEMP-s stars ([Ba/Fe] $> 1.0$, [Ba/Eu] $> 0$) and CEMP-no stars ([Ba/Fe] $< 1.0$).

The gray dashed line in the carbon panel of Figure 8 represents [C/Fe] = +0.7. As shown in Figure 8, nine stars are identified as CEMP stars. Since we mainly focus on the stars with clear origins, three of these CEMP stars (two from GSE and one from in situ) are marked with the open stars in Figure 8. All these three CEMP stars are CEMP-s stars, with [Ba/Fe] = 1.5 and 1.6 for GSE members and [Ba/Fe] = 1.75 for in situ stars, respectively. The chemical compositions of CEMP-s are changed due to mass transfer from their companions (C. Abate et al. 2015; T. T. Hansen et al. 2016). Therefore, we excluded them throughout the discussion of the chemical properties of our sample. We note that no CEMP-no stars are found in our GSE or in situ sample. The potential reason could come from two aspects: on the one hand, in the $-2.0 <$ [Fe/H] $< -1.0$ metallicity range, the proportion of CEMP stars is low ($\sim$10%, A. Arentsen et al. 2022); on the other hand, only 20 stars in our sample have carbon abundance measurement, primarily due to the detection limits of carbon in warmer stars (H. Li et al. 2022). Thus, the absence of CEMP-no stars in our sample is expected.

As shown in the oxygen panel, the oxygen abundances of GSE members are not significantly different from those of in situ stars. As metallicity increases, the abundances of O element begin to rise at [Fe/H] $> -1.2$. Compared to other elements, there are relatively fewer stars with carbon abundance or oxygen abundance measurements. This may be due to that C and O are relatively difficult to measure in warmer stars, such as MSTO stars.

### 3.3. α-elements

The abundances of α-elements were determined by the analysis of measured EWs. For all four measured α-elements,

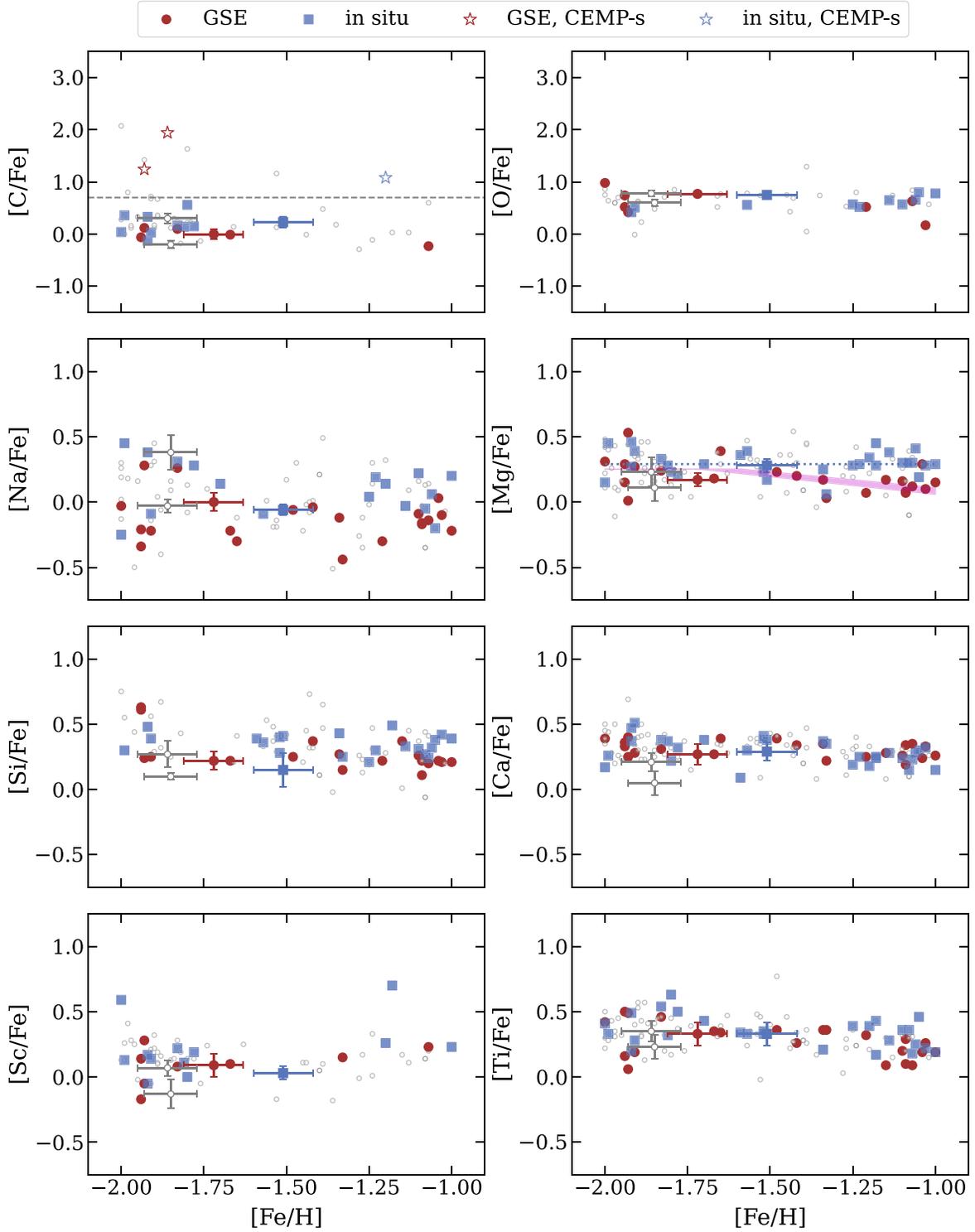

**Figure 8.** Abundance trend along metallicities for elements from carbon to titanium. The gray dashed line in the carbon panel is [C/Fe] = +0.7, representing the criterion of CEMP-s stars. The open star represents the CEMP-s stars. The red and blue lines in the Mg panel represent the evolution trend of Mg elements in GSE and in situ stars. The red shaded area marks the fitting uncertainty. The meanings of other symbols are the same as in Figure 7.

Mg, Si, Ca, and Ti, the number of stars with measurements in GSE is 27, 22, 27, and 28, and the number of stars with measurements in in situ components is 33, 24, 33, and 30. For [Si/Fe], there are relatively fewer stars with Si abundance measurements than other α-elements, which is attributed to the mixture of Si lines and the lack of spectral lines that can be used to derive Si abundances.

We show the distribution of the [α/Fe] versus [Fe/H] in Figure 8. In general, the Mg, Si, and Ti abundances in GSE stars show differences compared to in situ stars, while Ca abundances are largely similar. For all four measured α-elements, GSE and in situ stars show a similar enhancement of ∼0.3 at the lower metallicity region. However, GSE exhibits a significant decline in [Mg/Fe] at [Fe/H] > −1.6, whereas the [Mg/Fe] in in situ stars remain constant until [Fe/H] = −1.0. We fit the trend of [Mg/Fe] with increasing metallicity. The fitting results show that the position of $\alpha$-knee in GSE is around −1.60 ± 0.06, with a decreasing slope of −0.280 ± 0.013 at [Fe/H] > −1.6. For [Si/Fe] and [Ti/Fe], tentative offsets are observed within [Fe/H] > −1.6. The average Si and Ti abundances of GSE are both 0.24 at [Fe/H] > −1.6, while average Si and Ti abundances of in situ stars are 0.34 and 0.31, respectively.

The $\alpha$-knee of GSE we find is consistent with S. Monty et al. (2020, [Fe/H] ∼ −1.6) and the low-$\alpha$ sequence stars in P. E. Nissen & W. J. Schuster (2010, [Fe/H] ⩽ −1.5), slightly lower than the location identified by J. T. Mackereth et al. (2019) using APOGEE DR14 data ([Fe/H] = −1.3), and higher than that using all samples of the SAGA database ([Fe/H] ∼ −2.0; T. Matsuno et al. 2019). The difference in the position of $\alpha$-knee could be attributed to variations in analysis methods or the stellar types in the sample. The $\alpha$-knee position is a classic indicator to understand the star formation history of a galaxy (e.g., E. Tolstoy et al. 2009). A slower star formation rate, or an effectively truncated upper initial mass function, moves the knee to lower metallicities in (dwarf) galaxies with smaller masses. The $\alpha$-knee in the GSE occurs in lower metallicity than the Milky Way, indicating a less massive progenitor of GSE than the Galaxy, which is consistent with previous findings.

The distinction between the two components is less clear for [Ca/Fe] than the other three $\alpha$-elements, likely due to varying contributions from SNe Ia for different elements. According to T. Tsujimoto et al. (1995), the relative contribution of SNe Ia to the Ca abundance is greater than to the Mg abundance.

### 3.4. Light Odd-Z Elements

We derived abundances for two light odd-Z elements, Na and Sc. In our sample, Na was analyzed by LTE. We used the 5682 and 5688 Å lines to measure the Na abundances, which are less sensitive to NLTE departures compared to the resonant NaD lines at 5890/5896 Å which are usually used for VMP stars. These two lines can be easier measured for moderately metal-poor stars than VMP stars, thus greatly reducing the impact of the NLTE effect. The Na abundances from the SAGA database are LTE measurements (T. Suda et al. 2017), which is consistent with our analysis method. We used 14 Sc II lines to measure Sc abundances. Twenty-four GSE members and 20 in situ stars in our sample have Na measurements, while 11 GSE members and 13 in situ stars have Sc measurements.

The final [Na/Fe] are listed in Table 3 and shown in Figure 8. The [Na/Fe] exhibits a significant scatter, and, generally, GSE members show lower Na abundance than the in situ stars. The derived Sc abundance ratios of our sample are slightly overabundant compared to the solar value, with small scatters. The Sc abundances of GSE stars are not significantly different from those of in situ stars in the metallicity range of our sample.

### 3.5. Iron-peak Elements

Iron-peak elements are synthesized in SNe Ia events, during the final stages of the massive stars, as well as in incomplete or complete Si burning during explosive burning of core-collapse supernova (CCSNe; F. X. Timmes et al. 1995; S. E. Woosley & T. A. Weaver 1995; A. Chieffi et al. 1998; C. Kobayashi et al. 2006).

We measured the abundances of six more iron-group elements other than Fe, i.e., V, Cr, Mn, Co, Ni, and Zn. The V abundance has been determined from the V I line (4379 Å). Two Zn I lines at 4722 and 4810 Å were used to determine the abundance of Zn. Three lines have been used to derive abundances of Mn (4055, 4783, and 4823 Å) and Co (4092, 4118, and 4121 Å). We used quite a number of usable Cr I and Ni I lines to measure Cr and Ni. The number of stars in the sample with measurements of these elements is 10 (V), 22 (Cr), 18 (Mn), 9 (Co), 21 (Ni), and 17 (Zn) for the GSE, and 9 (V), 17 (Cr), 12 (Mn), 10 (Co), 18 (Ni), and 14 (Zn) for the in situ component.

The derived iron-group elemental abundances are shown in Figure 9 as a function of [Fe/H]. Chromium is produced by both CCSNe and SNe Ia. As seen in the [Cr/Fe]–[Fe/H] diagram, both GSE and in situ stars exhibit small scatter and display similar trends, indicating that there is no significant difference in [Cr/Fe] between GSE and in situ stars. Additionally, [Cr/Fe] shows an increasing trend with metallicity, consistent with previous studies (P. E. Nissen & W. J. Schuster 2011; R. Zhang et al. 2024).

P. E. Nissen & W. J. Schuster (2011) argued that accretion stars fall below the in situ components in the [Ni/Fe]–[Fe/H] diagram after the onset of SNe Ia. Our result shows that [Ni/Fe] decreases at [Fe/H] ∼ −1.6 in our GSE sample, which is consistent with this scenario. For the [Fe/H] > −1.6 subsample (nine GSE and seven in situ stars), the in situ population exhibits systematically higher [Zn/Fe] (with a median value of 0.17) and [Ni/Fe] (0.08) compared to GSE ([Zn/Fe], 0.08; [Ni/Fe], −0.07), though larger samples are needed to confirm this tentative trend. According to K. Iwamoto et al. (1999), SNe Ia produces relatively little Zn. Hence, the decreasing trend of [Zn/Fe] in the GSE may be explained by the onset of the noticeable contribution of SNe Ia. Meanwhile, the [Zn/Fe] ratios in in situ stars remain unchanged within this metallicity range.

Based on the available data, there is no significant difference between GSE and in situ stars in terms of V, Mn, and Co abundances.

### 3.6. Heavy Elements

The bulk of the chemical elements beyond the iron group are mainly synthesized through neutron-capture nuclear reactions since neutrons do not need to overcome the Coulomb barrier to penetrate into atomic nuclei. Depending on the timescales involved between neutron captures and $\beta$-decays of the radioactive nuclei, the mechanism can operate in two extreme regimes, which we call s-process and r-process (e.g., K. Nomoto et al. 2013). These two nucleosynthesis mechanisms require different astrophysical conditions.

We calculated the Sr abundances from two Sr resonance lines (4077 and 4215 Å). According to P. Hannaford et al. (1982), hyperfine structures in Y are insignificant. Thus, Y abundances were calculated from four Y lines (4374, 4854, 4883, and 4900 Å) without considering hyperfine structure. Zr abundances were obtained from two Zr lines (4161 and 4209 Å). The three heavier elements (Ba, La, and Eu) were determined from Ba II (4554, 4934, 6141, and 6497 Å), La II (4077, 4086, 4196, and 4921 Å), and Eu II (4129 and 4205 Å) lines. The number of stars in the sample with measurements of these elements in GSE is 12 (Sr), 22 (Y), 10 (Zr), 26 (Ba), 7 (La), and 13 (Eu), and in in situ is 18 (Sr), 20 (Y), 11 (Zr), 27 (Ba), 7 (La), and 8 (Eu).

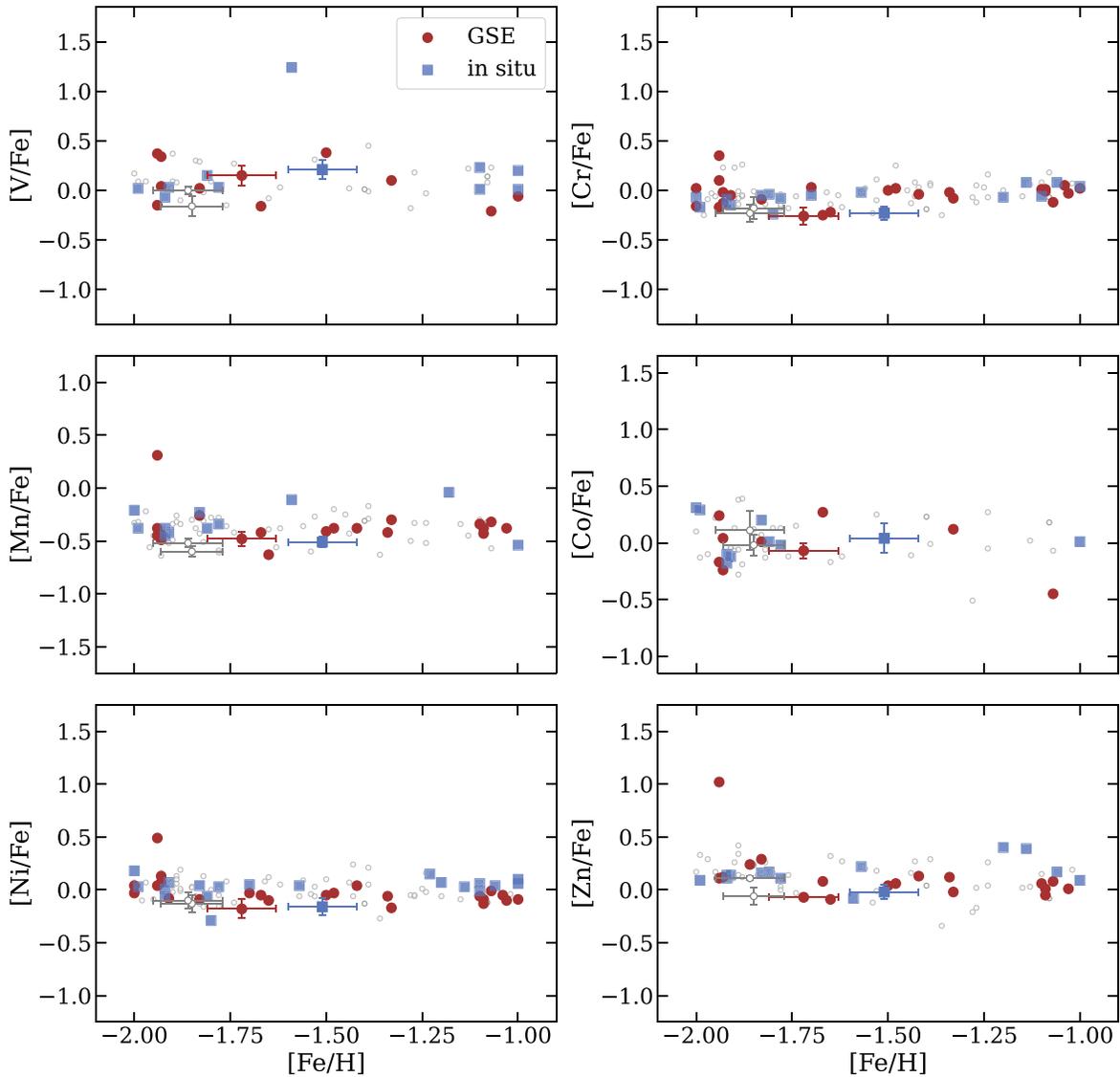

**Figure 9.** Abundance trend along metallicities for iron-group elements. The meanings of the symbols are the same as in Figure 7.

Figure 10 demonstrates the abundance trend for heavy elements. In the upper-left panel of Figure 10, the Sr abundance is close to the solar ratio. For both GSE and in situ components, Sr shows a similar trend as the metallicity increases. The upper-right panel of Figure 10 shows [Y/Fe] as a function of [Fe/H]. The GSE stars tend to have lower [Y/Fe] values than the in situ stars, and there is considerable scatter in [Y/Fe]. From the Galactic evolution model, C. Travaglio et al. (2004) predicted that both [Y/Fe] and [Ba/Fe] are close to zero in halo and thick-disk stars with [Fe/H] > −2.0. This is in rough agreement with the observed trend in Figure 10.

The lower-left panel of Figure 10 shows the abundance distribution of Ba. There is little difference between GSE and in situ stars. Most stars in our sample seem to group around [Ba/Fe] ∼ −0.2. The large scatter in Ba abundances within the metallicity range of our sample indicates a complex cosmic evolution, likely involving both the *s*-process and *r*-process (P. François et al. 2007; S. M. Andrievsky et al. 2009).

The lower-right panel of Figure 10 compares the observed Eu abundance ratios of our sample. Several studies have noted the enhancement in Eu of GSE (M. N. Ishigaki et al. 2013; D. S. Aguado et al. 2021; A. J. Koch-Hansen et al. 2021; T. Matsuno et al. 2021; A. Carrillo et al. 2022; S. Monty et al. 2024; X. Ou et al. 2024). In our GSE sample, a similar enhancement in Eu abundance ratio is found. The median [Eu/Fe] of GSE members is 0.55, which agrees with the literature that the Eu enhancement feature of GSE appears in the metallicity range of −2.0 < [Fe/H] < −0.7 (D. S. Aguado et al. 2021; T. Matsuno et al. 2021; A. Carrillo et al. 2022; X. Ou et al. 2024). So far, we cannot draw a clear conclusion on the trend of Eu abundance, since we do not have many stars with measured Eu abundance.

We present the distribution of [Ba/Y] and [Ba/Eu] in Figure 11. [Ba/Y] can reflect the delayed production of *s*-process elements by low-mass asymptotic giant branch stars (see, e.g., E. Tolstoy et al. 2009; P. E. Nissen & W. J. Schuster 2011). P. E. Nissen & W. J. Schuster (2011) show that [Ba/Y] of accretion compositions has an upward trend between the metallicity from −1.2 and −0.7. However, our sample only covers stars with $-2.0 <$ [Fe/H] $< -1.0$, so we cannot find a similar increase among GSE members. Besides, [Ba/Eu] is also an excellent indicator in the earlier phase to distinguish contributions of the r- and s-processes, since Eu is mainly produced by the r-process and Ba is synthesized via both r- and s- processes (see, e.g., T. Matsuno et al. 2021; X. Ou et al. 2024). In the early stage of the Universe, Ba was preliminarily produced by the r-process, and the s-process would gradually dominate the nucleosynthesis of Ba with the evolution of galaxies (C. Kobayashi et al. 2020). At [Fe/H] $\sim -2.0$, GSE members have lower [Ba/Eu] than in situ stars, which is due to the Eu enrichment in GSE. As the metallicity increases, we can clearly see an increasing trend in [Ba/Eu] of GSE members in Figure 11, implying that the contribution of the s-process to Ba increases with time in the progenitor of GSE.

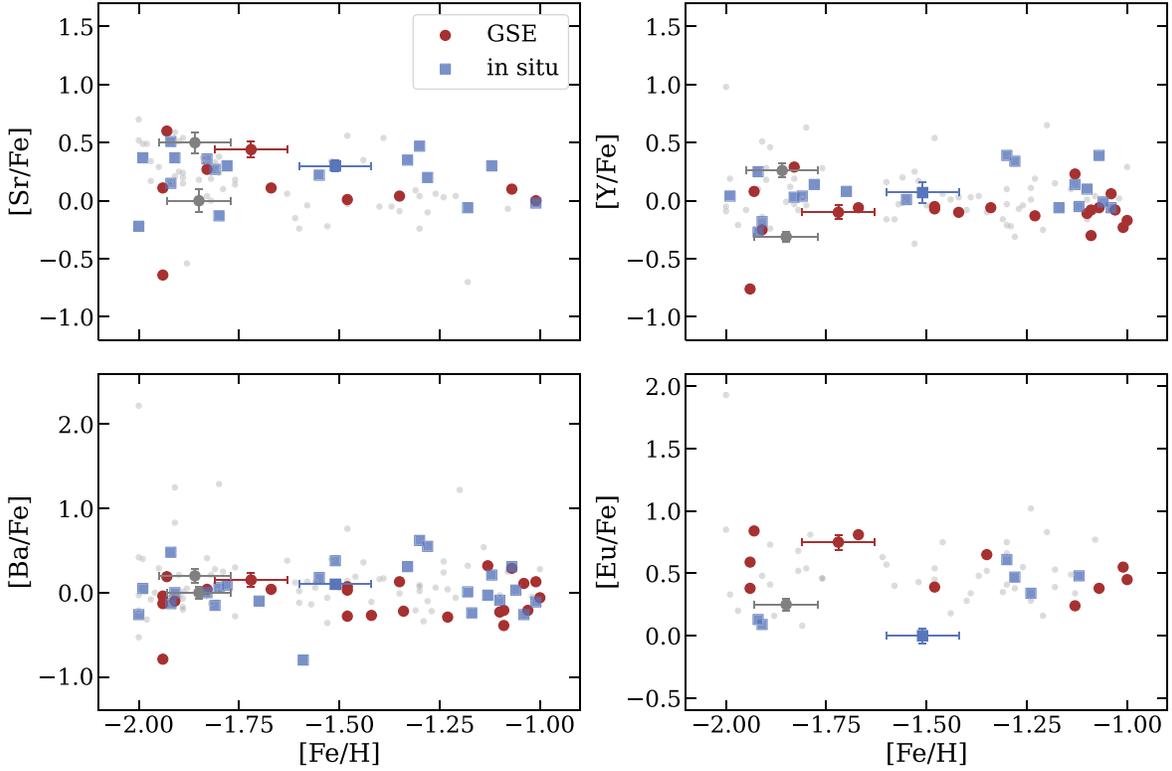

**Figure 10.** Abundance trend along metallicities for heavy elements. The meanings of the other symbols are the same as in Figure 7.

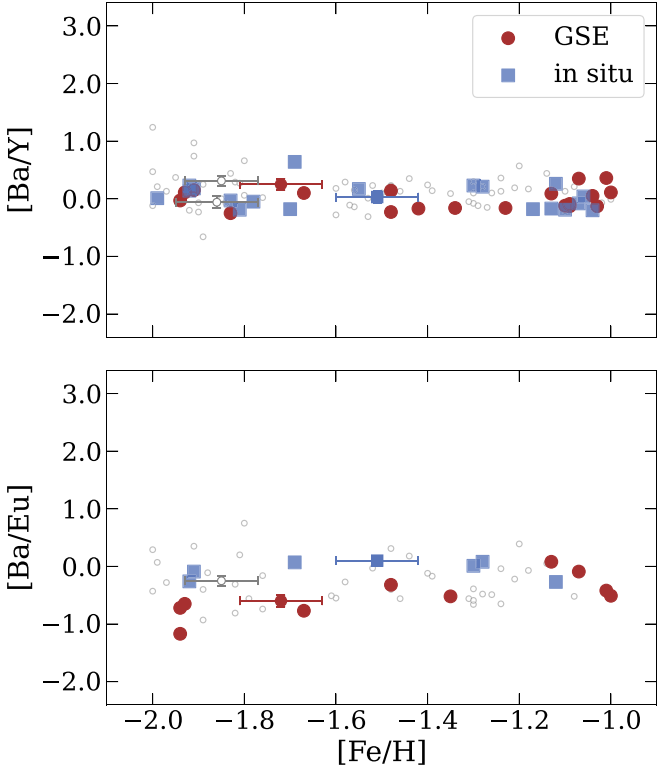

**Figure 11.** [Ba/Y] and [Ba/Eu] as a function of [Fe/H] with the same symbols as in Figure 10.

### 3.7. The Full r-process Pattern in GSE

The observed elemental abundance patterns of metal-poor r-process-enhanced stars can be used to characterize the yields of r-process production events. Evidence for stars displaying the r-process pattern, including the Sun and metal-poor stars, strongly suggests that the pattern is universal in the Milky Way, at least for elements from Ba to Hf (C. Sneden et al. 2003; K. Nomoto et al. 2013; A. Frebel et al. 2016; A. P. Ji & A. Frebel 2018). However, it remains uncertain whether the r-process pattern in dwarf galaxies (e.g., GSE) is similar to that of the Milky Way.

Previous studies have revealed the enhanced feature of r-process elements in GSE at [Fe/H] $> -2.0$ (D. S. Aguado et al. 2021; T. Matsuno et al. 2021; A. Carrillo et al. 2022). Since these studies only focus on a few heavy elements (e.g., Ba and Eu), studies on the full r-process pattern of GSE stars are very much needed to unveil the r-process nucleosynthesis in the GSE. The previous analysis shows that J0722 is an extremely r-process-enhanced GSE member (r-II stars with [Eu/Fe] $> 0.7$ and [Ba/Eu] $< 0$, D. Gudin et al. 2021). This star offers an opportunity to uncover the *r*-process pattern within GSE, thereby improving our comprehension of its nucleosynthetic history.

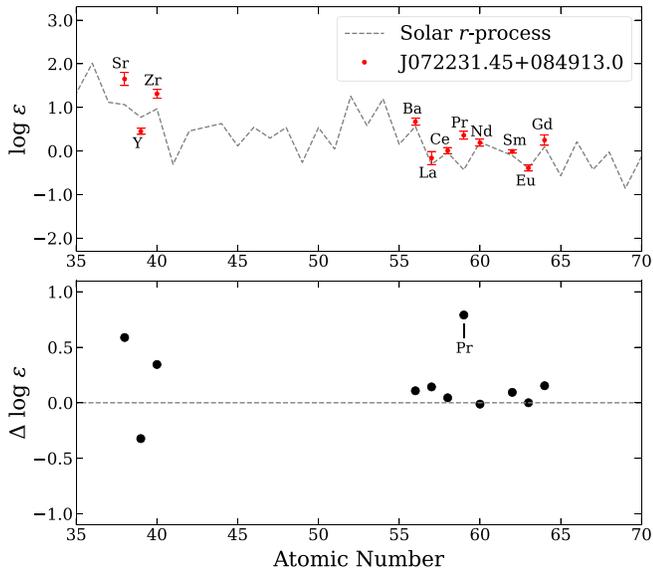

**Figure 12.** Upper panel: The red dots are the abundances of neutron-capture elements of J0722 with the error bars. The gray dashed line is the *r*-pattern of the solar system. The solar *r*-process pattern is shifted to match the Eu abundance. Lower panel: Abundance offsets between J0722 and the solar system.

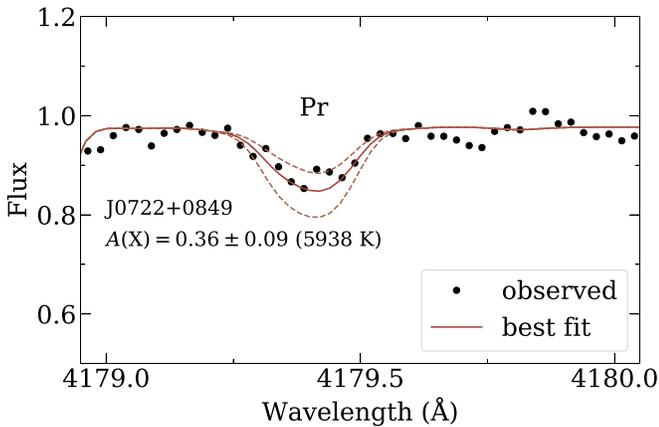

**Figure 13.** The spectral fitting result of Pr 4179 Å line for J0722 under LTE models. The meanings of the dots and lines are the same as in Figure 3. The best-fitting result of $A(X) = 0.36$ and uncertainty ($\pm 0.09$) are shown in solid and dashed lines, respectively.

To compare the *r*-process pattern of the identified GSE star J0722 with that of in situ, as shown in Figure 12, we plot the abundances of the neutron-capture elements for J0722 along with the solar *r*-process values, which is a representative in situ. It can be seen that J0722 is in good agreement with the *r*-process-enhanced model except for Pr. In order to justify the measurement of the element Pr, we show the measurement results of the Pr in Figure 13. It can be seen that the fitting of Pr and the derived abundance are reliable.

## 4. Conclusions

We conduct a detailed chemo-dynamical analysis on the moderately metal-poor MSTO stars to explore the early accretion history of the Milky Way. The MSTO stars are ideal tracers to study the chemical properties of different progenitor systems since their chemical abundances in the outer atmosphere are barely polluted by the mixing of internal materials.

Based on high-resolution optical spectroscopic data from CFHT, we derived atmospheric parameters and abundances for four MSTO stars. Additionally, we collected the elemental abundances from the SAGA database for 163 MSTO stars. We combined these stars as the final sample and calculated the dynamical parameters for the sample stars. Within the action-angle phase space, we identify a number of substructures, such as GSE (31), in situ (35), Thamnos (13), Sequoia (4), Helmi streams (3), and Wukong/LMS-1 (1). We use the moderately metal-poor MSTO stars to compare the detailed chemical abundance distribution between GSE members and the in situ stars. The major results include:

1. The Li abundance of both the GSE and in situ stars appears a plateau with a $A(Li) \sim 2.17$ at $-2.0 <$ [Fe/H] $< -1.0$. Moreover, the Li abundance and its scatter show no significant differences between the GSE and in situ stars in our sample, indicating that the star formation environment may not affect the formation of the Spite Plateau.

2. The $\alpha$-elements in GSE stars exhibit a similar enhancement $\sim 0.3$ as in situ stars toward the very low-metallicity region. However, we observe a very clear downward trend in Mg for GSE stars, the fitting results indicate that the $\alpha$-knee for the Mg element in GSE occurs at a metallicity of $-1.60 \pm 0.06$, with a slope of $-0.280 \pm 0.013$. For [Si/Fe] and [Ti/Fe], tentative decreasing trends are also observed within the same metallicity range, though these are less pronounced than that for Mg. In contrast, the $\alpha$-elements in in situ stars do not decrease until [Fe/H] $= -1.0$. The distinct [Mg/Fe] trend enables us to clearly distinguish between stars from the two stellar systems.

3. The iron-peak elements, except for Ni and Zn, show little difference between GSE and in situ stars. GSE stars exhibit a decreasing trend in Zn and Ni when it comes to [Fe/H] $> -1.6$. This indicates that the GSE began to experience SNe Ia contributions at a lower metallicity, suggesting that it is a relatively low-mass system.

4. Among heavy elements, Eu presents an enhancement in the GSE stars compared to the in situ stars, which is consistent with previous findings among giants. Moreover, a clear different trend can be detected for [Ba/Eu]. For the GSE sample stars, we see an increasing [Ba/Eu] along with increasing metallicity, whereas the ratio stays almost constant for in situ stars within the metallicity region of our sample. The increasing trend in [Ba/Eu] of GSE members implies that the contribution of the *s*-process to Ba increases with time in the progenitor of GSE.

5. For the first time, we derive the *r*-process abundance pattern for a GSE MSTO star, which is in general similar to the solar *r*-process pattern. However, an over $3\sigma$ deviation is noticed for Pr, which may require further exploration.

The criteria to define different substructures are debated, since different accretion events may overlap in dynamical phase space, and a single accretion event may lead to multiple components with different kinematic characteristics (I. Jean-Baptiste et al. 2017; R. P. Naidu et al. 2021; J. A. S. Amarante et al. 2022; G. Pagnini et al. 2023). Therefore, obtaining detailed chemical information is crucial for truly understanding the origins of the Galactic substructures, since the outer atmosphere of metal-poor stars preserve the chemical imprint of their progenitors. As can also be seen in this work, the chemical information of moderately metal-poor MSTO stars in the literature is still insufficient to support systematic analysis of substructures much smaller than the GSE. Furthermore, for several elements (e.g., C, O, and heavy elements), the measurements for MSTO stars are even insufficient for systematic discussion of the majority of inner halo (in situ stars and the GSE). Our future work will analyze detailed chemical abundances for MSTO stars in more substructures and, thereby, the chemo-dynamical properties of a larger MSTO sample.

The dynamical picture of the Milky Way has greatly evolved thanks to the spectacular view from Gaia, while the chemo-dynamical picture of the Milky Way is just emerging in the upcoming era of spectroscopic surveys (e.g., SDSS-V, WEAVE, PFS, and 4MOST). Undoubtedly, the combination of detailed chemical abundances and dynamics will provide better evidence for testing our models of the formation and accretion history of the Milky Way and its satellites. The next step is to expand the sample and analyze the chemical abundance of more member stars in the specific substructures.

## Acknowledgments

This work is supported by the Strategic Priority Research Program of Chinese Academy of Sciences grant No. XDB1160100, the National Natural Science Foundation of China grant Nos. 12222305 and 12588202, the National Key *R&D* Program of China Nos. 2024YFA1611903 and 2023YFE0107800, the International Partnership Program of Chinese Academy of Sciences (grant No. 178GJHZ2022040GC), and the China Manned Space Program with grant No. CMS-CSST-2025-A12.

Guoshoujing Telescope (the Large Sky Area Multi-Object Fiber Spectroscopic Telescope, LAMOST) is a National Major Scientific Project built by the Chinese Academy of Sciences. Funding for the project has been provided by the National Development and Reform Commission. It is operated and managed by the National Astronomical Observatories, Chinese Academy of Sciences. This research uses data obtained through the Telescope Access Program (TAP), which has been funded by the National Astronomical Observatories of China, the Chinese Academy of Sciences (the Strategic Priority Research Program "The Emergence of Cosmological Structures" Grant No. XDB09000000), and the Special Fund for Astronomy from the from the Ministry of Finance. This work is based on observations obtained at the Canada–France–Hawaii Telescope (CFHT) which is operated by the National Research Council of Canada, the Institut National des Sciences de l'Univers of the Centre National de la Recherche Scientifique of France, and the University of Hawaii. This work has made use of data from the European Space Agency (ESA) mission Gaia (https://www.cosmos.esa.int/gaia), processed by the Gaia Data Processing and Analysis Consortium (DPAC, https://www.cosmos.esa.int/web/gaia/dpac/consortium). Funding for the DPAC has been provided by national institutions, in particular the institutions participating in the Gaia Multilateral Agreement. This research has used the SAGA database (T. Suda et al. 2008).

## ORCID iDs

Renjing Xie https://orcid.org/0009-0002-4282-668X
Haining Li https://orcid.org/0000-0002-0389-9264
Ruizhi Zhang https://orcid.org/0009-0008-1319-1084
Yin Wu https://orcid.org/0009-0003-4116-6824
Xiang-Xiang Xue https://orcid.org/0000-0002-0642-5689
Gang Zhao https://orcid.org/0000-0002-8980-945X
Xiao-Jin Xie https://orcid.org/0000-0002-4440-4803